\documentclass[pra,twocolumn,superscriptaddress]{revtex4}
\usepackage{epsfig}
\usepackage{amsmath}
\usepackage{amsfonts}
\usepackage{amssymb}
\usepackage{mathrsfs}
\usepackage{color}

\usepackage{times}

\usepackage{ifpdf}
\ifpdf
\usepackage{epstopdf}   
\fi

\newcommand{\ket}[1]{\ensuremath{\vert#1\rangle}}
\newcommand{\bra}[1]{\ensuremath{\langle #1\vert}}
\newcommand{\bk}[2]{\ensuremath{\langle #1\vert #2\rangle}}
\newcommand{\kb}[2]{\ensuremath{\vert #1 \rangle \langle #2 \vert}}

\renewcommand{\vec}[1]{\ensuremath{\mathbf{#1}}}

\def\unity{\mbox{\small 1} \!\! \mbox{1}}

\begin{document}

\title{Measurement based entanglement under conditions of extreme photon loss}

\author{Earl T. Campbell}
%\email{earl.campbell@materials.ox.ac.uk}
\affiliation{Department of Materials, University of Oxford, Parks Road, Oxford OX1 3PH, UK.}
\author{Simon C. Benjamin}
\email{s.benjamin@qubit.org}
\affiliation{Department of Materials, University of Oxford, Parks Road, Oxford OX1 3PH, UK.}
\affiliation{Centre for Quantum Technologies, National Univeristy of Singapore, 3 Science Drive 2, Singapore 117543.}

\begin{abstract}

The act of measuring optical emissions from two remote qubits can entangle them. By demanding that a photon from each qubit reaches the detectors, one can ensure than no photon was lost. But the failure rate then rises quadratically with loss probability. In~\cite{MMOYMDM1a} this resulted in 30 successes per billion attempts. 
We describe a means to exploit the low grade entanglement heralded by the detection of a lone photon: A subsequent perfect operation is quickly achieved by consuming this noisy resource. We require only two qubits per node, and can tolerate both path length variation and loss asymmetry. The impact of photon loss upon the failure rate is then linear; realistic high-loss devices can gain orders of magnitude in performance and thus support QIP. 
\end{abstract}

\maketitle 

\begin{figure}[t]
\centering
\includegraphics[width=7.8cm]{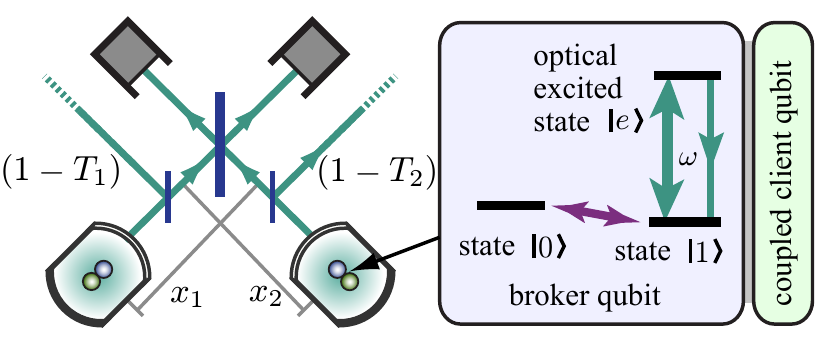}
\caption{Schematic of an apparatus suitable for creating entanglement between distant entities, using a beam splitter for erasure of {\em which path} information. Our model incorporates uncertain optical path length and asymmetric photon loss via the parameters shown.  Photon loss is modelled by beam splitters with transmittance $T_{1}$ and $T_{2}$; loss at different parts of the apparatus has an equivalent description in terms of $T_{1}$ and $T_{2}$.  We define $\phi$ to be the transmittance asymmetry, such that $\sin(2 \phi)=(T_{1}-T_{2})/(T_{1}+T_{2})$.  Asymmetry in path length is parametrised by $\Delta= \pi (x_{1}-x_{2})/\lambda$.
}
\label{fig:apparatusAndProtocol}
\end{figure}

Recent experimental successes~\cite{MMOYMDM1a, MMOYMM01a, BJDDMBG01a,newMoehring} demonstrate that measuring two quantum systems can cause them to become entangled. Consider an optically active matter system: an atom, or an atom-like structure such as a quantum dot or a crystal defect. Following laser stimulation such a system may emit a photon; its internal state is then correlated with the photon. We may generate entanglement by stimulating two separate systems simultaneously and monitoring their emissions in such a way that we determine characteristics of their mutual state {\em without} learning the source of any given photon~\cite{CCFZ01a} (e.g. Fig.~\ref{fig:apparatusAndProtocol}). This approach, and certain comparable alternatives~\cite{LMNSLBM01a, HYOR01a}, have the profound advantage that the component systems can be far apart, which may make a {\em scalable} technology more attainable. 

Any photon loss from the apparatus will leave the matter systems in an uncertain state. Unfortunately real systems will always suffer from finite photon capture efficiency, and this can be exasperated by non-ideal physics in emission process. For example nitrogen-vacancy (NV) defects in diamond emit less than three percent of their light cleanly, i.e. without giving rise to phonons. A scalable technology would need further lossy components for optical routing. Photon loss is therefore to be recognised as a fundamental difficulty.

One solution is {\em weak excitation}: reduce the rate of photon generation so that emission from both qubits simultaneously is very rare. A detector click is taken to imply that exactly one photon was emitted, and successfully reached the detectors~\cite{CCFZ01a,BKPV01a, BPH01a}. However the rate of entanglement generation is then inversely linked to the fidelity. Entanglement between atomic ensembles has been accomplished in this way~\cite{CRFPEK01a}. In alternative {\em two-photon} schemes both matter systems emit a photon, and success is heralded by detecting them both~\cite{DK01a, FZLGX01a, SI01a, BK01a, LBK01a, BES01a}. We register a failure if either photon is lost, so the success rate falls as the square of the transmission probability. Recently two macroscopically separate atoms were entangled in this way~\cite{MMOYMDM1a} at a success rate of about thirty per billion attempts (with a recent preprint reporting a 13 fold improvement~\cite{newMoehring}).

Here we describe a new approach to entanglement generation that can be fundamentally more rapid. Adopting an entanglement distillation technique, we develop a protocol that can quickly implement a perfect gate operation using the imperfect entanglement that results from seeing a lone detector `click'. Our gate operation is a parity projection, which supports universal quantum computing through the graph state approach; we evaluate the performance in this context. The approach requires two qubits at each local site. This modest level of complexity has already been demonstrated in several systems: In atom trap devices one can confine two atoms, while recent experiments on NV defects have shown the interacting electron and nuclear spins are individually controllable~\cite{DuttScience07}. The utility of such two-qubit nodes has been studied in the context of quantum repeaters~\cite{quantRep}. 

In order to provide a clear exposition, we will assume the specific energy level structure shown in Fig.~\ref{fig:apparatusAndProtocol}. Alternatives such as $\Lambda$ structures can be equally suitable. We begin by preparing each of the optically active qubits, which term {\em brokers} following Ref.~\cite{BBFM01a}, in state 
\begin{equation}
\ket{\theta}= \cos (\theta) \ket{0} + \sin (\theta) \ket{e}.
\end{equation}
Here $\vert e \rangle$ is the state that decays radiatively to $\ket{1}$, see Fig.~\ref{fig:apparatusAndProtocol}. We collect photons using a lens or a cavity system and direct them through a beam splitter.  Given an ideal apparatus, if exactly one photon is detected then the brokers are projected onto the $(\ket{01}+\ket{10})/\sqrt2$ state (neglecting any phase that depends on \textit{which} detector clicked, which is trivially corrected by a local operation).  However in reality we may have non-number resolving detectors, photon loss (which may be asymmetric) and path length variations.  Hence, a single click corresponds to a mixed state
\begin{equation}
\rho_{B} =(1-\eta) \mathcal{Z}^{\phi, \Delta}_{B1} \vert \Psi^+ \rangle \langle \Psi^{+} \vert \mathcal{Z}^{\phi, -\Delta}_{B1} \\ +\eta \vert 11 \rangle \langle 11 \vert ,
\label{stateETA}
\end{equation}
where operator $\mathcal{Z}^{\phi, \Delta}_{B1}$ represents the effects of asymmetry in the apparatus, as parameterized in Fig.~\ref{fig:apparatusAndProtocol}. This operator can be formally represented as acting on broker qubit $B1$ alone:
\begin{equation}
	\mathcal{Z}^{\phi, \Delta}_{B1} = [\cos (\phi) \unity + \sin (\phi) Z_{B1}][\cos (\Delta) \unity + i \sin (\Delta) Z_{B1}].
\label{ZopEqn}
\end{equation}
In Appendix~\ref{sec:eta} we show that the probability that we indeed see a click, i.e. that we obtain $\rho_B$ from initial state 
$\ket{\theta}\ket{\theta}$, is
\begin{equation}
P_{click} = T \sin^2(\theta)\left( 2 - T \sin^2(\theta) \cos^{2}(2 \phi) \right)
\label{pClickEqn}
\end{equation}
where $T=(T_{1}+T_{2})/2$. The weighting in $\rho_B$ is shown to be
\begin{equation}
\eta = \frac{\sin^{2} (\theta)[2-T \cos^{2}( 2 \phi ) ]}{2-T \sin^{2} (\theta) \cos^{2} (2 \phi) }.
\end{equation}
  
From the work of Bennett~
{\em et al.}~\cite{BDSW01a} we know that there is a finite chance of distilling a perfect Bell state from any two states of the form $(1-\eta)\vert \Psi^+ \rangle \langle \Psi^{+} \vert +\eta \vert 11 \rangle \langle 11 \vert$. Note this is equivalent to (\ref{stateETA}) without imperfections $\Delta$ and $\phi$.  Here we show that one can perform a perfect {\em parity projection} by consuming two (or more) resource states of the general form  (\ref{stateETA}). We find that $\Delta$ and $\phi$ can be completely unknown provided that they do not drift significantly over the course of the protocol, and that two qubits per node suffice for all operations.  Thus we obtain the high fidelity typical of two photon schemes, but at a rate that is affected only linearly by $T$.  

We present our protocol in terms of a basic {\em iterate} that must be performed more than once. The aim is to perform a single high fidelity parity projection on two remote {\em client} qubits.
Note that although we are adopting the {\em broker-client} terminology from Ref.~\cite{BBFM01a}, that earlier paper did not exploit the brokers for distillation and thus implicitly assumed that the initial entanglement mechanism is near perfect ($\eta\rightarrow 0$ and $\mathcal{Z}\rightarrow\unity$ in Eqn.(\ref{stateETA})). The first flow chart shows one iterate:

\begin{figure}[!h]
\centering
\includegraphics[width=8.cm]{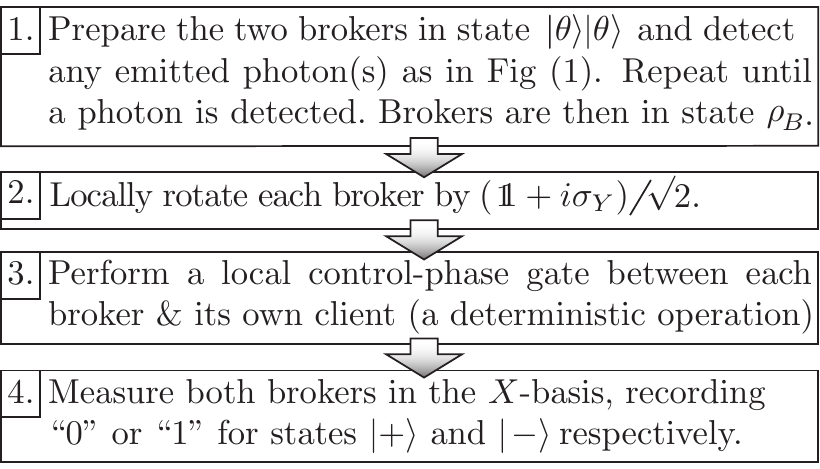}
\label{fig:loop}
\end{figure}
Step 2 will map $\ket{11} \rightarrow \ket{++}$ and $\mathcal{Z}^{\phi, \Delta}_{B1} \ket{\Psi^{+}} \rightarrow \mathcal{X}^{-\phi, -\Delta}_{B1} \ket{\Phi^{-}}$, where $\ket{\Phi^-}=(\ket{00}-\ket{11})/\sqrt{2}$ and
\[
	\mathcal{X}^{\phi, \Delta}_{B1} = [\cos (\phi) \unity + \sin (\phi) X_{B1}][\cos (\Delta) \unity + i \sin (\Delta) X_{B1}].
\]
Using $i$ and $j$ to denote measurement results on the brokers, one complete iterate will transform clients qubits from an initial state $\rho_{C}$ to an (unnormalised) final state (see Appendix~\ref{sec:QuantOp}):
\begin{equation}
\label{eqn:Eij}
E_{i,j}(\rho_{C}) \equiv\frac{1}{2}(1-\eta) \mathcal{Z}_{C1}^{\phi_{i},\Delta_{i}} P_{i,j} \rho_{C} P_{i,j} \mathcal{Z}_{C1}^{\phi_{i},-\Delta_{i}}  + \eta S_{i,j} \rho_{C} S_{i,j},
\end{equation}
where $P_{i,j}$ is a parity projector with parity opposite to that of $i,j$, and $S_{i,j}$ is an undesired projector onto separable states. Specifically $S_{i,j}=\kb{i,j}{i,j}$ while $P_{i,j}= \left( \unity - (-1)^{i+j} Z_{C1}Z_{C2} \right) /2$, the labels $C1$ and $C2$ referring to the two client qubits.  The variables $\phi_{i}$ and $\Delta_{i}$ are $-(-1)^{i}\phi$ and $-(-1)^{i}\Delta$, respectively. Since this state is mixed, we will require least one more iterate. Initially we will consider two iterates in total, moving to more general cases later.

\begin{figure}[!h]
\centering
\includegraphics[width=8.2cm]{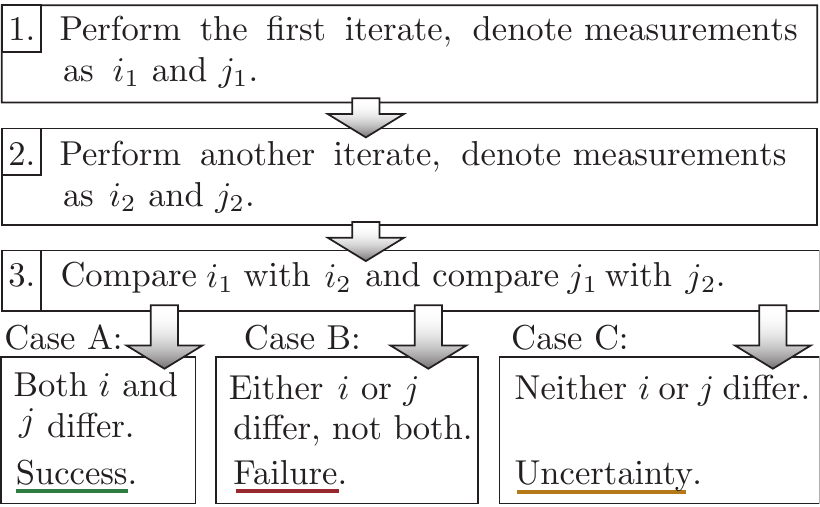}
\label{fig:flow_protocol}
\end{figure}

In Case A the we have measurement results of the same parity, but differing specific $i$, $j$ values.  The corresponding parity projectors are consistent, $P_{i_n,j_n}P_{i_1,j_1} = P_{i_1,j_1}$, but all other projector combinations vanish, $P_{i_n,j_n}S_{i_1,j_1} = S_{i_n,j_n}P_{i_1,j_1} = S_{i_n,j_n}S_{i_1,j_1} = 0$. We conclude with certainty that the clients qubits have simply been acted on by  $\mathcal{Z}_{C1}^{\phi_{i},\Delta_{i}} \mathcal{Z}_{C1}^{-\phi_{i},-\Delta_{i}} P_{i_1,j_1}$, which is proportional to a pure parity projection, as $\mathcal{Z}_{C1}^{\phi_{i},\Delta_{i}} \mathcal{Z}_{C1}^{-\phi_{i},-\Delta_{i}}=\cos (2 \phi) \unity$.  The probability of reaching Case A after two iterations is
\begin{equation}
\label{eqn:Ptwo}
P_{two}= \cos^{2}(2 \phi)(1-\eta)^{2}/2 .
\end{equation}
In Case B the parity of $i$,$j$ has changed and consequently $P_{i_n,j_n}P_{i_1,j_1} =0$, so we must conclude that the client qubits have been projected into a separable state. This is an unrecoverable failure and the clients must be reset before trying again (any prior entanglement with other qubits is lost).  In Case C the quantum operation on the clients still contains some asymmetry operator, $(\mathcal{Z}_{C1}^{\pm \phi_{i}, \pm \Delta_{i}})^{m}$, and will also produce a mixture if all $i$,$j$ have been the same. We could choose to abort here, declaring Case C a failure, which would be a \textit{2 iterates only} ($2IO$) strategy. Later we will show that one can also persist with further iterations in order to resolve the uncertainty.

\begin{figure}[t]
\centering
\includegraphics[width=8cm]{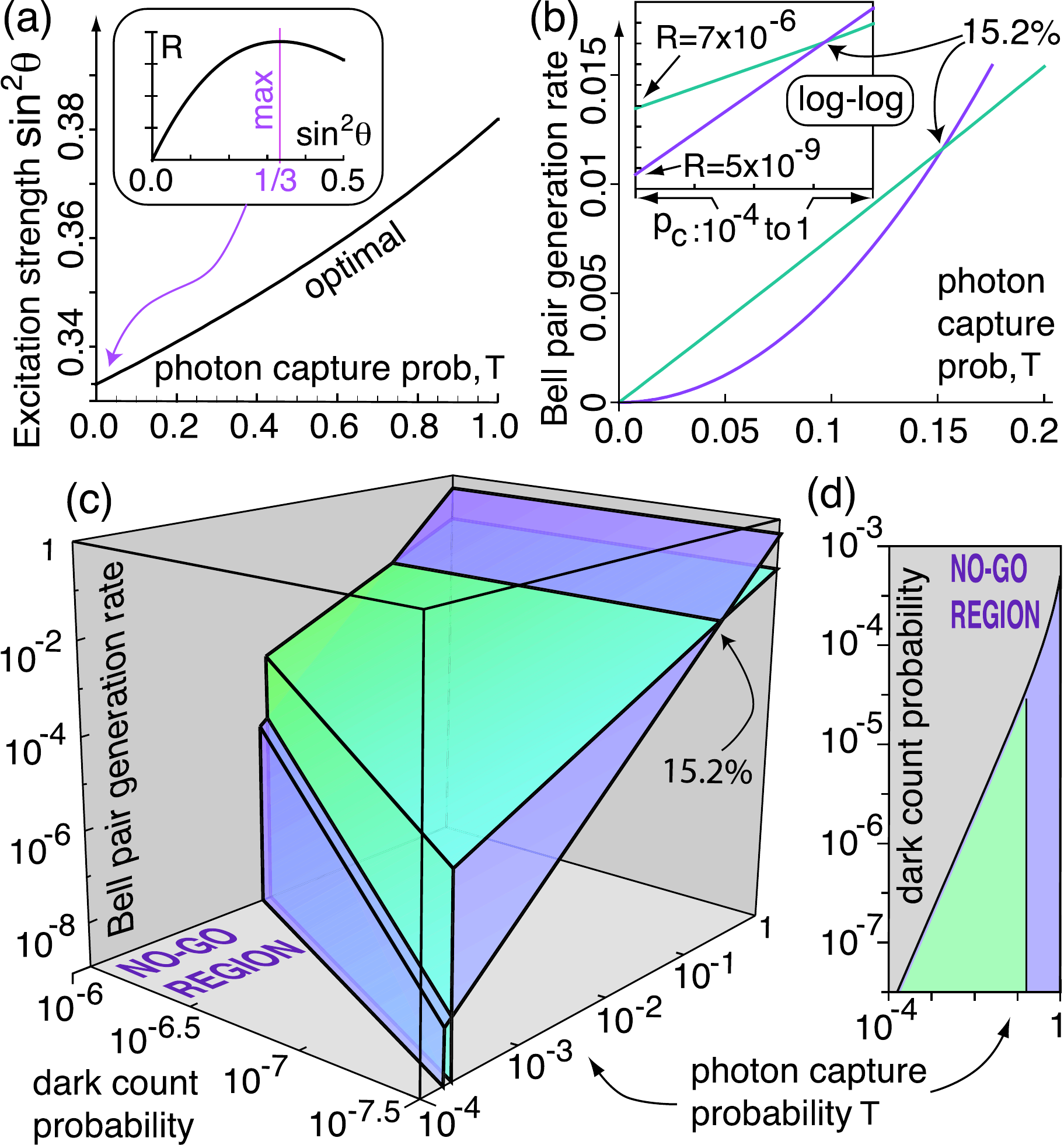}
\caption{Attainable rate of Bell pair production, $R$. (a) Optimising experimental parameter $\theta$ (corresponding to amplitude of $\ket{e}$ in initial state $\ket{\theta}$) to maximise $R$. (b) Linear and log-log (inset) plots comparing the rate of Bell pair production in units of $1/\tau$ using our approach (green) versus a comparable prior scheme, i.e. the double heralding protocol~\cite{BK01a} (blue). Other schemes involving detecting two-photons \cite{DK01a, FZLGX01a, SI01a, BK01a, LBK01a} will have similar performance in the high photon loss domain. Note the crossover point of $15.2\%$ above which our approach is unhelpful. (c) Regions in which our approach (green) and double-heralding (blue) have superior rates of generation, given a finite probability of a dark count occurring while we monitor for photons following excitation of the matter systems. Here we assume a minimum acceptable fidelity of $1-10^{-3}$, thus there is a `no-go' parameter region where both approaches fail. (d) A top-down view, over a wider range of dark counts.  For simplicity we assume here that photon loss is symmetric: $T_1=T_2=T$.}
\label{fig:plots}
\end{figure}

In Fig.~\ref{fig:plots} we characterise the performance in the simplest scenario: the $2IO$ protocol acting on clients initialised to $\ket{+}\ket{+}$, thus producing Bell pairs at a rate $R$.  The rate is expressed in units of $1/\tau$, where $\tau$ is the time for a single attempt at generating a photon. Rate $R$ is optimised by a specific $\theta$ as shown in Fig.~\ref{fig:plots}(a) and detailed in Appendix~\ref{sec:EPR_prod}. In Fig.~\ref{fig:plots}(b) we compare our approach to a {\em two photon} entanglement scheme, finding three orders of magnitude improvement when $T$ reaches $10^{-4}$ (the approximate value that has been achieved experimentally~\cite{MMOYMDM1a}).

\begin{figure}[t]
\centering
\includegraphics[width=8cm]{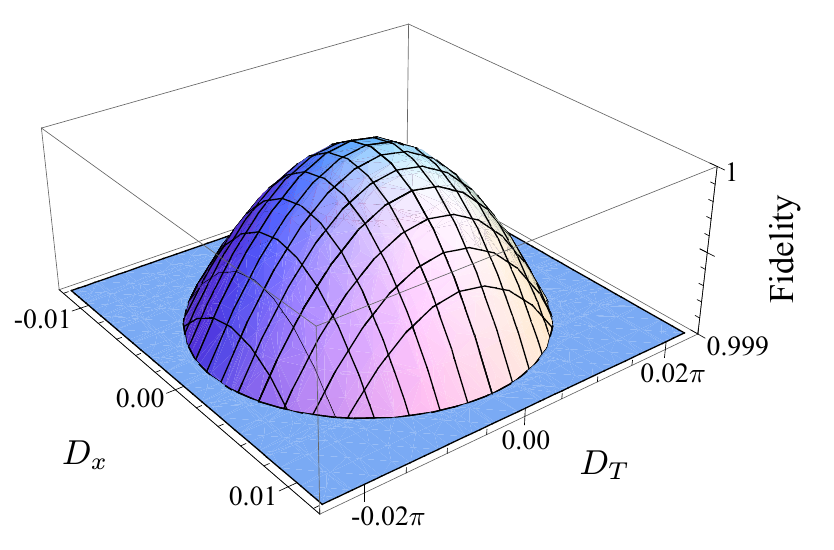}
\caption{The fidelity of the $2IO$ strategy in the presence of drifting apparatus cut off at fidelity $1-10^{-3}$.  Drift in the path length is quantified by $D_x=((x_{1}-x_{2})-(x'_{1}-x'_{2}))/\lambda$, where unprimed (primed) variables represent the initial (final) values.  Drift in the photon transmission probabilities is quantified by $D_T=1-\sqrt{(T'_{1}T_{2})/(T_{1}T'_{2})}$, with the same priming notation.}
\label{fig:jitter_final}
\end{figure}

Dark counts can be a primary cause of infidelity in entanglement achieved by the path-erasure approach~\cite{MMOYMDM1a}. The approach described here, with its single broker-client pair at each location, is vulnerable to accumulation of dark count noise over successive iterations. However because the number of iterations is so small, the accumulation need not be substantial. In Fig.~\ref{fig:plots}(c) we make the same comparison as in \ref{fig:plots}(b) but now with finite dark counts and insisting on a minimum fidelity of $1-10^{-3}$. The thresholds defining the `no-go' region are nearly identical: the reference scheme can tolerate marginally higher dark counts, about $ 5\%$ greater, and still achieve the target fidelity (note the narrow shelf just visible in the 3D plot). Note that to enter the no-go region one could introduce further ancilla qubits and combine the protocol described here with schemes like Refs.~\cite{DB01a, JTSL02a, duan04, ODH01a}.

We have shown that our protocol does not require one to determine any difference in path length between the `arms' of the device, nor any asymmetry  in the transmission probabilities; the former will cancel and the later only reduces the overall probability of success without degrading the entanglement. However, this conclusion is only valid if the unknown quantities do not drift significantly {\em during} a given instance of the protocol. If indeed such drift has occurred, but is relatively slight, then it is straightforward to characterise the level of resulting error. In Appendix~\ref{sec:drift} we perform this calculation and find that the infidelity
$\epsilon =  (\pi D_{x} )^{2} + (D_{T}/2)^{2} + {\rm Order}(D_x,D_T)^4$. In Fig.~\ref{fig:jitter_final} we show the fidelity of the parity projection of a $2IO$ strategy, where drift occures between the first and second iterate. 
Assuming that only path length drift is significant, the infidelity will  be below $10^{-3}$ provided that drift $D_x<1/(32 \pi)$.  

\begin{figure}[t]
\centering
\includegraphics[width =8.3cm]{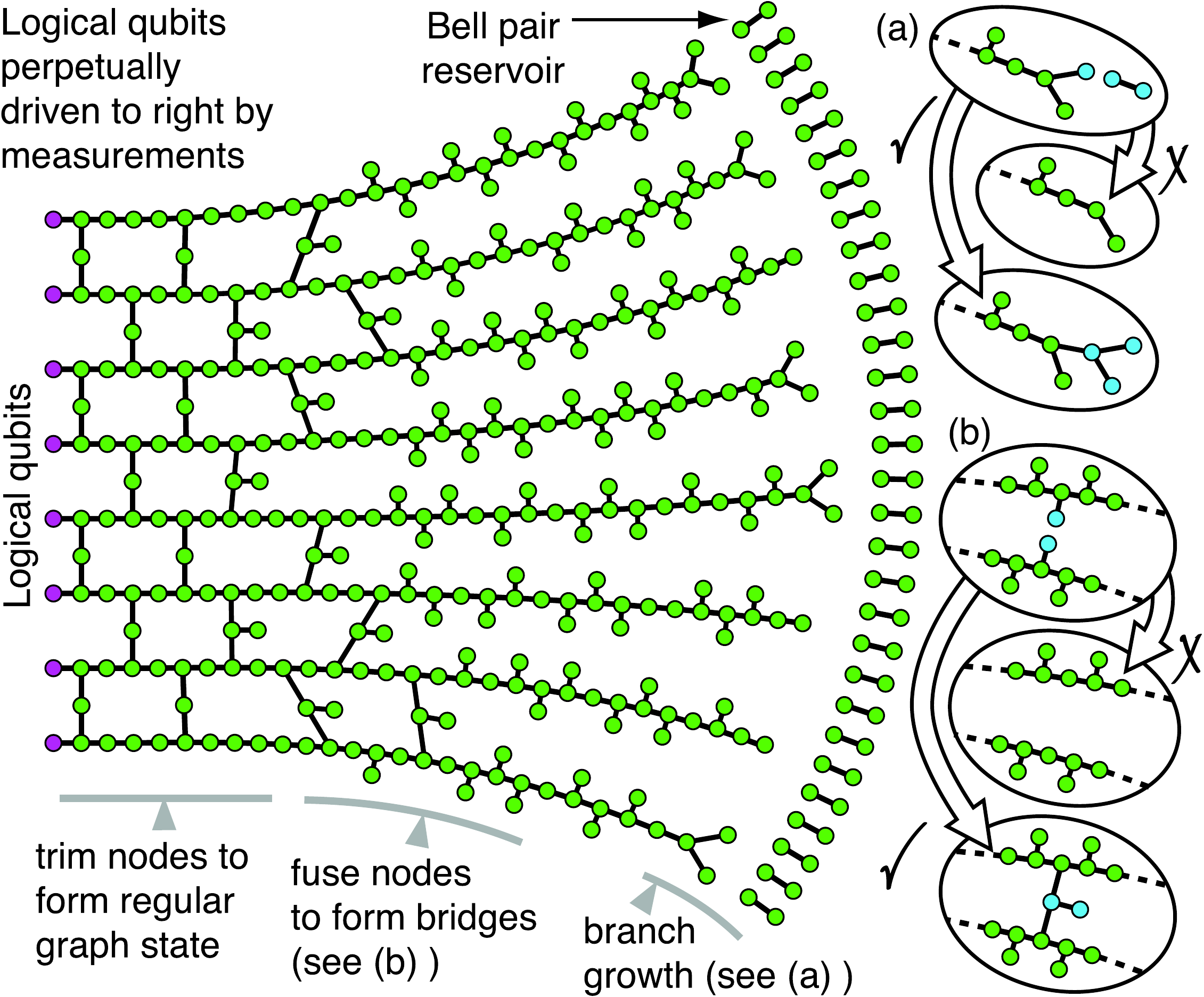}
\caption{A suitable method for full scale quantum computation using the well studied {\em graph state} entanglement. The approach depicted here provides the context for us to evaluate our protocol. Each dot (or `node') represents a qubit, each line (or `edge') represents entanglement between qubits. On the left side the graph state has a regular ordering, using which one can perform a computation purely by measuring out individual qubits (each in an appropriate basis)~\cite{NDMB01a,RBB01a, HDERNB01a}. The graph is thus consumed from the left, driving the logical qubits to the right such that they are forever on the left-most fringe. Simultaneously a second set of measurements, i.e. those corresponding to the entanglement process described here, generate fresh graph state structure (right edge) and fuse it into a regular array (central). Entanglement is created `just in time' for use. As each physical system is measured out on the left, it is reused in the process of new growth. The small figures to the right show the results of success (creation of new entanglement) or failure (destruction of prior entanglement). }
\label{fig:graph}
\end{figure}

Our protocol has applications far beyond simply generating high quality Bell pairs. Because we perform a parity projection on the client qubits rather than purifying a specific state, we can use the same protocol to subsequently entangle those qubits with other partners. It is now well understood that one can thus build up a many-qubit resource such as a graph state, and so support efficient universal quantum computation~\cite{NDMB01a,RBB01a, HDERNB01a}. We will evaluate the performance of our protocol in the context of the specific growth strategy depicted in Fig.~\ref{fig:graph}. There are now two numbers to optimise, the rate of new entanglement generation, and the probability that a given entanglement operation will succeed. Thus it is interesting to generalise our simple {\em 2 iterates only} protocol in order that `Case C' need not be abandoned. In fact this is straightforward: Over multiple iterates, outright failure occurs only when the parity of measurement outcomes $i$, $j$ in the latest iterate differs from the previously seen parity.  Success requires that all outcomes have the same parity {\em and} the two possible instances of that parity have occurred the {\em same number of times}. The latter criterion assures that any errors described by $\mathcal{Z}$ will not degrade the resulting entanglement; our former expression for the state projector simply generalises to $(\mathcal{Z}_{C1}^{\phi_{i},\Delta_{i}})^{\frac{n}{2}} (\mathcal{Z}_{C1}^{-\phi_{i},-\Delta_{i}})^{\frac{n}{2}} P_{i_1,j_1}$, which reduces to a pure parity projection as before. Given a system where $\mathcal{Z}\rightarrow\unity$, this criterion could of course be relaxed.

In the paradigm illustrated in Fig.~\ref{fig:graph}, one would employ the simple {\em 2IO} strategy for the creation of the Bell pair reservoir, and the more sophisticated multi-iterate protocol for the creation of other `edges' where failure will cause damage (as shown in the Fig.~\ref{fig:graph} insets). The characteristic time required to perform operations on the logical qubits, i.e. the effective clock speed of the quantum computer, depends on the time needed to create new edges in the graph. It is straightforward to determine this variable in the limit of low photon capture probability $T$; the calculation appears in Appendix~\ref{sec:Chain_Growth}. The average time required to increase a branch length by one qubit, accounting for both successes and failures of the form shown in Fig.~\ref{fig:graph}(a), is found to be $29.0(\tau/T)$. This is in contrast to ${\rm Order}(\tau / T^2)$ for two-photon schemes such as Ref.~\cite{BK01a}. We emphasise that although this `clock speed' depends on $T$, in our approach the probability of success is substantially {\em de-coupled} from $T$. Finally we note that although this example is 2D, one can equivalently generate higher dimensional graphs, including the structures suggested by Raussendorf {\em et al.} for fault tolerant computation with a high threshold \cite{RHG01a}. 

We have described a form of measurement-based entanglement generation that can be used in situations of high photon loss. Performance can be orders of magnitude higher than current experiments, thus bringing fault tolerance thresholds correspondingly closer and enhancing the prospects for a highly scalable technology. We thank Joe Fitzsimons, Sean Barrett, Jason Smith, Pieter Kok and David Moehring for helpful comments. This research was supported by the Royal Society and the QIP IRC.

\appendix

\section{A derivation of $\eta$ \& $P_{click}$.}
\label{sec:eta}

We denote $P_{click}$ as the probability of single detector click from two brokers in the state $\ket{\theta}\ket{\theta}$.  This is simply the sum of: the probability of a single photon being emitted and captured,
\begin{equation}
	P_{(1)} = 2 T \sin^{2} ( \theta ) \cos^{2} ( \theta );
\end{equation}
and the probability of two photons being emitted and \textit{either} being captured
\begin{equation}
 P_{(2)} = \sin^{4}( \theta ) [ 1 - ( 1- T )^{2} ] ; 
\end{equation}
Note that the \textit{either} is inclusive since photons bunch and detector are assumed to be non-number resolving.  The resulting mixed state has a $\ket{11} \bra{11}$ contribution with a proportion $\eta=P_{(2)}/P_{click}$.

\section{The quantum operation $E_{i,j}(\rho_{C})$.}
\label{sec:QuantOp} 

We derive the quantum operation for measurement results $i$ \& $j$ by considering how each pure state $\ket{\Psi^{+}}$ and $\ket{11}$ affects the clients, and then weighting these outcomes appropriately.  Considering $ \mathcal{Z}_{C1}^{\phi, \Delta}   \ket{\Psi^{+}}$ first, the local rotation $( \unity + i Y )/\sqrt{2}$ maps this to the state $\mathcal{X}_{B1}^{-\phi,-\Delta}  \ket{\Phi^{-}}$. Next a control-Z is performed between broker and client, followed by an $X-$basis measurement, resulting in:
\begin{equation}
	\bra{\pm}_{B} CZ_{B}^{C} = ( \bra{0}_{B} \otimes \unity_{C} \pm \bra{1}_{B} \otimes Z_{C} )/ \sqrt{2}.
\end{equation}
It is easy to verify that this operator maps the asymmetry operators between brokers and clients, such that:
\begin{equation}
	\bra{\pm}_{B1} CZ_{B1}^{C1} \mathcal{X}_{B1}^{-\phi,-\Delta} = \bra{\pm}_{B1} CZ_{B1}^{C1} \mathcal{Z}_{C1}^{ \mp \phi, \mp \Delta}.
\end{equation}
Using $i \& j$ to represent the measurement outcomes on brokers $B1$ and $B2$, we conclude that the resulting effect on the client qubits is:
\begin{eqnarray}
&& \bra{(-1)^{i}, (-1)^{j}}_{B1, B2} CZ_{B1}^{C1} CZ_{B1}^{C1} \mathcal{X}_{C1}^{-\phi,-\Delta}  \ket{\Phi^{-}} \\ \nonumber
& = & \bra{(-1)^{i}, (-1)^{j}}_{B1, B2} CZ_{B1}^{C1} CZ_{B1}^{C1}   \ket{\Phi^{-}} \mathcal{Z}_{B1}^{ \phi_{i}, \Delta_{i} } \\ \nonumber
& = & (\unity_{C1}  \unity_{C2} - (-1)^{i+j} Z_{C1} Z_{C2})  \mathcal{Z}_{B1}^{ \phi_{i}, \Delta_{i} } /2\sqrt{2} \\ \nonumber
& = & P_{i,j}  \mathcal{Z}_{C1}^{ \phi_{i}, \Delta_{i} } / \sqrt{2}
\end{eqnarray}
where $\phi_{i}= -(-1)^{i} \phi$ and similarly $\Delta_{i}= -(-1)^{i} \Delta$.  Hence, the $\ket{\Psi^{+}}$ component projects the clients into a parity subspace that is the opposite of the measurement outcome, which is a projector that we denote $P_{i,j}$.  

When we consider the $\ket{11}$ component, its separability means that it is sufficient to derive the effect of a single broker on its clients, and the same will hold for the second broker.  For a single broker the rotation $(\unity + i Y)/ \sqrt{2}$ maps $\ket{1} \rightarrow \ket{+}$, which is followed by a control-Z operation to give:
\begin{equation}
 (\ket{0} + \ket{1} Z_{C})/\sqrt{2}
\end{equation}
so measuring a broker in the state $\ket{+}(\ket{-})$ means that its client is projected into the separable state $\ket{0} (\ket{1})$.  Taking both brokers in consideration, the measurement signature $i,j$, results in the projector $S_{i,j}= \ket{i,j} \bra{i,j}$.  

To construct a quantum operation we begin by considering that $\ket{\Psi^{+}}$ and $\ket{11}$ contributions will occur with probability $(1- \eta)$ and $\eta$.  However, the occurrence of a parity projector is shared between two different measurement signatures, which introduces a factor of $1/2$.  This gives an (unnormalised) quantum operation of:
\begin{equation}
	E_{i,j}(\rho_{C}) = \frac{1}{2}(1-\eta) P_{i,j} \rho_{C} P_{i,j} + \eta S_{i,j} \rho_{C} S_{i,j}.
\end{equation}
Note that the probability of each outcome can be calculated by taking the trace of $E_{i,j}(\rho_{C})$.  As a check it is reassuring to calculate $\sum_{i,j \in \{ 0,1 \} } \mathrm{tr}[ E_{i,j}(\rho_{C})] =1$.

\section{Drift effects}
\label{sec:drift}

Since $\mathcal{Z}^{\phi, \Delta} \mathcal{Z}^{{-\phi, -\Delta}} \propto \unity$, we know that for two distillation iterates with different parity outcomes asymmetry operators vanish.  Of course, this assumes that the asymmetry is constant over the time between two iterates.  In this section we consider the errors caused by drift or jitter in the apparatus:
\begin{eqnarray}
	 \epsilon & = & 1 -  \frac{| \bk{\Psi^{+}}{\Psi_{\rm{drift}}}   |^{2}}{\bk{\Psi_{\rm{drift}}}{\Psi_{\rm{drift}}}} ,
\end{eqnarray}
where,
\begin{eqnarray}
	\ket{\Psi_{\rm{drift}}} = \mathcal{Z}^{\phi + \delta \phi, \Delta + \delta \Delta} \mathcal{Z}^{{-\phi, -\Delta}} \ket{\Psi^{+}}
\end{eqnarray}
where $\delta \phi$ and $\delta \Delta$ quantify the apparatus drift between first and second iterate.  This expands out to a function of asymmetry variables $\phi$ \& $\Delta$, and their associated drifts $\delta \phi$ \& $\delta \Delta$:
\begin{eqnarray}
	 \epsilon & = &  \frac{\cos^{2} (2 \phi + \delta \phi ) \sin^{2} (\delta \Delta) + \sin^{2} (\delta \phi) \cos^{2} (\delta \Delta ) } { \cos ^{2} (2 \phi + \delta \phi )  + \sin^{2} (\delta \phi)}.
\end{eqnarray}
This equation does not directly lend itself to an interpretation in terms of physical properties of the apparatus, so we introduce new variables:
\begin{eqnarray}
	D_{x} & = & \left( (x_{1} - x_{2}) - (x'_{1} - x'_{2}) \right)/ \lambda \\ \nonumber
	D_{T}  & = & 1 - \sqrt{  \frac{T'_{1} T_{2}}{T'_{2} T_{1}} } 
\end{eqnarray}
where unprimed (primed) physical parameters describe the apparatus for the first (second) iterate of distillation.  In terms of these variables, we find:
\begin{eqnarray}
	\epsilon & = &  \frac{\sin^{2}(\pi D_{x}) + \cos^{2}(\pi D_{x})\left( \frac{D_{T}}{2+D_{T}}  \right)^{2} }{1+ \left( \frac{D_{T}}{2+D_{T}}  \right)^{2}}.
\end{eqnarray}
Expanding this expression to leading order in $D_{x}$ and $D_{T}$, we have a small-error approximation that is quadratic in both variables:
\begin{eqnarray}
	\epsilon & = & (\pi D_{x})^{2} + (D_{T}/2)^{2} + O[D_{x}, D_{T}]^{4}
\end{eqnarray}

\section{Rate of EPR production}
\label{sec:EPR_prod}

In this section we detail how to calculate the rate that Bell pairs can be produced.  In calculating this rate we have the preparation angle $\theta$ as a free parameter that we may optimise over.  This varies as a function of the overall photon loss, and also the asymmetry in photon loss.  

Using the $2IO$ strategy, the rate of Bell pair production is:
\begin{equation}
	 R  = \frac{1}{2} \cdot P_{\mathrm{two}} \cdot P_{click} / \tau,
\end{equation}
where $P_{two}$ is the probability of succeeding after 2 iterations, and $P_{click}$ is probability of succeeding at making the noisy Bell pairs required to drive the distillation.  The factor of $1/2$ is present since 2 iterations are required for each attempt.  Finally, the variable $\tau$ is the time taken for each attempt at the optical protocol used to make noisy Bell pairs (i.e. each attempt within step (1) of the iterate). We neglect the time taken to perform the local operations used in the distillation protocol, since we are principally interested in the regime of heavy photon loss where the dominant time cost comes from the many attempts at obtaining $\rho_B$.  Expressions for the probabilities, $P_{two}$ and $P_{click}$, can be found in the main paper, and using these we can obtain the production rate:
\begin{equation}
	 R   =  \frac{ T  \cos^{2}(2 \phi) \sin^{2} ( \theta )   \cos^{4} ( \theta )}{ 2 - T \sin^{2}( \theta ) \cos^{2} (2 \phi) }  / \tau.
\end{equation}
It is easy to show that in the limit of vanishing $T \cos^{2}(2 \phi) $, the optimal of our free parameter is $\sin^{2}(\theta)=1/3$.   For comparison, the strong excitation used in comparable {\em two photon} schemes would correspond to $\sin^{2}(\theta)=1/2$.  Since we are primarily interested robustness against photon loss, the figures provided in the main paper are optimal assuming that there is no asymmetry $\phi=0$, and that the amount of photon loss is known.  If these quantities are not known then the effect on rate will only be slight.

\section{Rate of chain growth}
\label{sec:Chain_Growth}

In this section we show how to reproduce the optimal growth rate when using a loop strategy to add Bell pairs to a chain.  The essential difference between Bell pair growth and chain growth, is that the latter requires consideration of how distillation failures reduce chain length.  This consideration means that a loop strategy gives a superior efficiency, as it avoids failure at any cost.   The rate at which qubits are added to a chain is:
\begin{equation}
	 G =  \frac{( 2 P_{loop} - (1- P_{loop}) ) P_{\mathrm{click}} }{ \langle I \rangle \tau } ,
\end{equation}
where the new variable $P_{loop}$ and $\langle I \rangle$, respectively represent: the probability of success without having to reset the client qubits; and the expected number of iterates until a distillation attempt is concluded by a success or reset.  The two terms in the numerator of $G$ are the success and failure contributions to chain length.  Since we are primarily concerned with robustness against photon loss, we will neglect the effects of photon loss asymmetry.

Before calculating the success probability, $P_{loop}$, and expected number of iterates until an attempt is concluded, $\langle I \rangle$, we must first find the various probabilities for every $k^{th}$ iterate.  Denoting $k^{th}$ iterate failure and success probabilities by $P_{f}(k)$ and $P_{s}(k)$, respectively, it follows that:
\begin{eqnarray}
	P_{loop} & = & \sum_{k=2}^{\infty} P_{s} (k), \\ \nonumber
	\langle I \rangle & = & \sum_{k=2}^{\infty} k  \left( P_{s} (k) + P_{f} (k) \right) .
\end{eqnarray}
Both success and failure on a $k^{th}$ iterate, may occur via a number of different sequences of measurement results, which must be summed over, so:
\begin{eqnarray}
	P_{f}(k) & = &   \left[ 2 \eta \left(   \frac{1- \eta}{2} \right)^{k-1} + \eta(1- \eta)^{k-1} \right]  \\ \nonumber
					 & + & 2 N_{f} (k) \eta \left( \frac{1-\eta}{2} \right)^{k-1} ,
\end{eqnarray}
and,
\begin{eqnarray}
P_{s}(k) & = & 2 N_{s}(k) \left( \frac{1- \eta}{2} \right)^{k}. 
\end{eqnarray}
The failure outcome has two terms representing: the probability of failing after a sequence of prior iterates with identical measurement results; and the second represents the all other possible combinations of prior measurement results, of which there are $N_{f}(k)$.  These terms differ because in the first instance, the density matrix is mixed prior to failure, whereas in the latter case the density matrix is pure prior to failure.  As for the probability of success, this again depends on the number of different sequences of measurement results, $N_{s}(k)$, which will obviously only be non-zero for even values of $k$.

Lastly, we must find a method of calculating the different valid combinations of measurement results, $N_{s}(k)$ and $N_{f}(k)$.  We begin by defining a vector $\vec{v}(k)$ which describes number of different $k-1$ measurement sequences prior to failure.  Prior to failure the all measurement results will be of the same parity, but this leaves two possible outcomes.  Using the variable $i$ to denote the difference in the number of the two possible outcomes, the $(i+1)^{th}$ entry of $\vec{v}(k)$ is the corresponding number of possible sequences with $i$ different measurements.  
 
Having defined $\vec{v}(k)$, we calculate it iteratively, by using:
\begin{equation}
	\vec{v}(k)  =  M^{k-2} \vec{v}(2) ,
\end{equation}
where $M$ has non-zero elements everywhere except above and below the diagonal, where the elements are $1$; Formally, $M_{i,j} = \delta_{i, j \pm 1}$.  The vector $\vec{v}(2)$, at which the first success can occur, has only the first entry as unity and the rest are zero.  In matrix form, we have:
\begin{equation}
	\vec{v}(k)  =  \left( \begin{array}{c c c c c c}
	 0 & 1 & 0 & \ldots & 0 & 0 \\
	 1 & 0 & 1 & \ldots & 0 & 0 \\
	 0 & 1 & 0 & \ldots & 0 & 0 \\
	 0 & 0 & 1 & \ldots & 0 & 0 \\
	 \vdots & \vdots & \vdots & \ddots & \vdots & \vdots \\
	 0 & 0 & 0 & \ldots & 1 & 0 \\
	\end{array} \right)^{k-2} \left( \begin{array}{c}
		1 \\
		0	\\
		0 \\
		. \\
		:	\\
		0	\\
	\end{array} \right).
\end{equation}

Now $N_{s}(k)$ and $N_{f}(k)$ follow directly, as the former is simply the first component, $N_{s}(k)= \vec{v}_{1}(k)$, and the failure variable is a sum from $i=2$ to the penultimate non-zeo entry, $N_{f}(k)= \sum_{i=2}^{k-1} \vec{v}_{i}(k)$.  

We have outlined all the necessary steps required to calculate $G$, and doing this for different values of $\eta$ will  allow use to find the optimal growth rate.  Performing this calculation for small photon capture, $T$, and negligible capture asymmetry, one finds that the optimal $G$ is $G=0.0345 T / \tau$.  The reciprocal of this rate, $29.0 \tau / T$, gives the average time taken to extend the chain by one qubit.


\begin{thebibliography}{25}

\expandafter\ifx\csname natexlab\endcsname\relax\def\natexlab#1{#1}\fi
\expandafter\ifx\csname bibnamefont\endcsname\relax
  \def\bibnamefont#1{#1}\fi
\expandafter\ifx\csname bibfnamefont\endcsname\relax
  \def\bibfnamefont#1{#1}\fi
\expandafter\ifx\csname citenamefont\endcsname\relax
  \def\citenamefont#1{#1}\fi
\expandafter\ifx\csname url\endcsname\relax
  \def\url#1{\texttt{#1}}\fi
\expandafter\ifx\csname urlprefix\endcsname\relax\def\urlprefix{URL }\fi
\providecommand{\bibinfo}[2]{#2}
\providecommand{\eprint}[2][]{\url{#2}}

\bibitem[{\citenamefont{Moehring et~al}(2007)\citenamefont{Moehring, Maunz,
  Olmschenk, Younge, Matsukevich, Duan, and Monroe}}]{MMOYMDM1a}
\bibinfo{author}{\bibfnamefont{D.~L.} \bibnamefont{Moehring et~al}},
  \bibinfo{journal}{Nature} \textbf{\bibinfo{volume}{449}}, \bibinfo{pages}{68}
  (\bibinfo{year}{2007}).

\bibitem[{\citenamefont{Maunz et~al}(2007)\citenamefont{Maunz, Moehring,
  Olmschenk, Younge, Matsukevich, and Monroe}}]{MMOYMM01a}
\bibinfo{author}{\bibfnamefont{P.}~\bibnamefont{Maunz et~al}},
  \bibinfo{journal}{Nature Physics} \textbf{\bibinfo{volume}{3}},
  \bibinfo{pages}{538} (\bibinfo{year}{2007}).

\bibitem[{\citenamefont{Beugnon et~al}(2006)\citenamefont{Beugnon, Jones,
  Dingjan, Darqui\'{e}, Messin, Browaeys, and Grangier}}]{BJDDMBG01a}
\bibinfo{author}{\bibfnamefont{J.}~\bibnamefont{Beugnon et~al}},
  \bibinfo{journal}{Nature} \textbf{\bibinfo{volume}{440}},
  \bibinfo{pages}{779} (\bibinfo{year}{2006}).
  
 \bibitem{newMoehring}D. N. Matsukevich {\em et al}, preprint arXiv:0801.2184v1.

\bibitem[{\citenamefont{Cabrillo et~al}(1999)\citenamefont{Cabrillo, Cirac,
  Garc?a-Fern?ndez1, and Zoller}}]{CCFZ01a}
\bibinfo{author}{\bibfnamefont{C.}~\bibnamefont{Cabrillo et~al}},
  \bibinfo{journal}{Phys. Rev. A} \textbf{\bibinfo{volume}{59}},
  \bibinfo{pages}{1025} (\bibinfo{year}{1999}).

\bibitem[{\citenamefont{Bose et~al}(1999)\citenamefont{Bose, Knight, Plenio,
  and Vedral}}]{BKPV01a}
\bibinfo{author}{\bibfnamefont{S.}~\bibnamefont{Bose et~al}},
  \bibinfo{journal}{Phys. Rev. Lett} \textbf{\bibinfo{volume}{83}},
  \bibinfo{pages}{5158} (\bibinfo{year}{1999}).
  
  \bibitem[{\citenamefont{Browne et~al}(2003)\citenamefont{Browne, Plenio, and
  Huelga}}]{BPH01a}
\bibinfo{author}{\bibfnamefont{D.~E.} \bibnamefont{Browne}},
  \bibinfo{author}{\bibfnamefont{M.~B.} \bibnamefont{Plenio}},
  \bibnamefont{and} \bibinfo{author}{\bibfnamefont{S.~F.}
  \bibnamefont{Huelga}}, \bibinfo{journal}{Phys. Rev. Lett}
  \textbf{\bibinfo{volume}{91}}, \bibinfo{pages}{067901}
  (\bibinfo{year}{2003}).
  
\bibitem[{\citenamefont{van Loock et~al}(2007)\citenamefont{van Loock, Munro,
  Nemoto, Spiller, Ladd, Braunstein, and Milburn}}]{LMNSLBM01a}
\bibinfo{author}{\bibfnamefont{P.}~\bibnamefont{van Loock et~al}},
 (\bibinfo{year}{2007}),
  \eprint{arXiv:quant-ph/0701057}.

\bibitem[{\citenamefont{Hu et~al}(2007)\citenamefont{Hu, Young, O'Brien, and
  Rarity}}]{HYOR01a}
\bibinfo{author}{\bibfnamefont{C.~Y.} \bibnamefont{Hu et~al}},
  (\bibinfo{year}{2007}), \eprint{arXiv:0708.2019}.

\bibitem[{\citenamefont{Chou et~al}(2005)\citenamefont{Chou, de~Riedmatten,
  Felinto, Polyakov, van Enk, , and Kimble}}]{CRFPEK01a}
\bibinfo{author}{\bibfnamefont{C.~W.} \bibnamefont{Chou et~al}},
  \bibinfo{journal}{Nature}
  \textbf{\bibinfo{volume}{438}}, \bibinfo{pages}{828} (\bibinfo{year}{2005}).

\bibitem[{\citenamefont{Duan and Kimble}(2003)}]{DK01a}
\bibinfo{author}{\bibfnamefont{L.~M.} \bibnamefont{Duan}} \bibnamefont{and}
  \bibinfo{author}{\bibfnamefont{H.~J.} \bibnamefont{Kimble}},
  \bibinfo{journal}{Phys. Rev. Lett.} \textbf{\bibinfo{volume}{90}},
  \bibinfo{pages}{253601} (\bibinfo{year}{2003}).

\bibitem[{\citenamefont{Feng et~al}(2003)\citenamefont{Feng, Zhang, Li, Gong,
  and Xu}}]{FZLGX01a}
\bibinfo{author}{\bibfnamefont{X.~L.} \bibnamefont{Feng et~al}},
  \bibinfo{journal}{Phys. Rev. Lett} \textbf{\bibinfo{volume}{90}},
  \bibinfo{pages}{217902} (\bibinfo{year}{2003}).

\bibitem[{\citenamefont{Simon and Irvine}(2003)}]{SI01a}
\bibinfo{author}{\bibfnamefont{C.}~\bibnamefont{Simon}} \bibnamefont{and}
  \bibinfo{author}{\bibfnamefont{W.~T.} \bibnamefont{Irvine}},
  \bibinfo{journal}{Phys. Rev. A} \textbf{\bibinfo{volume}{91}},
  \bibinfo{pages}{110405} (\bibinfo{year}{2003}).

\bibitem[{\citenamefont{Barrett and Kok}(2005)}]{BK01a}
\bibinfo{author}{\bibfnamefont{S.~D.} \bibnamefont{Barrett}} \bibnamefont{and}
  \bibinfo{author}{\bibfnamefont{P.}~\bibnamefont{Kok}},
  \bibinfo{journal}{Phys. Rev. A} \textbf{\bibinfo{volume}{71}},
  \bibinfo{pages}{060310} (\bibinfo{year}{2005}).

\bibitem[{\citenamefont{Lim et~al}(2005)\citenamefont{Lim, Beige, and
  Kwek}}]{LBK01a}
\bibinfo{author}{\bibfnamefont{Y.~L.} \bibnamefont{Lim}},
  \bibinfo{author}{\bibfnamefont{A.}~\bibnamefont{Beige}}, \bibnamefont{and}
  \bibinfo{author}{\bibfnamefont{L.~C.} \bibnamefont{Kwek}},
  \bibinfo{journal}{Phys. Rev. Lett} \textbf{\bibinfo{volume}{95}},
  \bibinfo{pages}{030505} (\bibinfo{year}{2005}).

\bibitem[{\citenamefont{Benjamin et~al}(2005)\citenamefont{Benjamin, Eisert,
  and Stace}}]{BES01a}
\bibinfo{author}{\bibfnamefont{S.~C.} \bibnamefont{Benjamin}},
  \bibinfo{author}{\bibfnamefont{J.}~\bibnamefont{Eisert}}, \bibnamefont{and}
  \bibinfo{author}{\bibfnamefont{T.~M.} \bibnamefont{Stace}},
  \bibinfo{journal}{NJP} \textbf{\bibinfo{volume}{7}},
  \bibinfo{pages}{194} (\bibinfo{year}{2005}).
  
\bibitem[{\citenamefont{Dutt et~al}(2007)\citenamefont{Dutt,  Childress, 
Jiang, Togan, Maze, Jelezko, Zibrov, Lukin}}]{DuttScience07}
\bibinfo{author}{\bibfnamefont{M. V. G.}~\bibnamefont{Dutt et~al}},
  \bibinfo{journal}{Science} \textbf{\bibinfo{volume}{316}},
  \bibinfo{pages}{1312} (\bibinfo{year}{2007}).
  
  \bibitem{quantRep}L. Childress {\em et al},
Phys. Rev. A, {\bf 72}, 052330 (2005).

\bibitem{JTSL02a} Jiang {\em et~al}, Phys. Rev. A {\bf 76}, 062323 (2007).

\bibitem[{\citenamefont{Benjamin et~al}(2006)\citenamefont{Benjamin, Browne,
  Fitzsimons, and Morton}}]{BBFM01a}
\bibinfo{author}{\bibfnamefont{S.~C.} \bibnamefont{Benjamin et~al}},
  \bibinfo{journal}{New Journal of Physics}
  \textbf{\bibinfo{volume}{8}}, \bibinfo{pages}{141} (\bibinfo{year}{2006}).
  
  \bibitem[{\citenamefont{Duan et~al.}(2004)\citenamefont{Duan, Blinov, Moehring,
  and Monroe}}]{DBMM01a}
\bibinfo{author}{\bibfnamefont{L.~M.} \bibnamefont{Duan}},
  \bibinfo{author}{\bibfnamefont{B.~B.} \bibnamefont{Blinov}},
  \bibinfo{author}{\bibfnamefont{D.~L.} \bibnamefont{Moehring}},
  \bibnamefont{and} \bibinfo{author}{\bibfnamefont{C.}~\bibnamefont{Monroe}},
  \bibinfo{journal}{Quantum Information and Computation}
  \textbf{\bibinfo{volume}{4}}, \bibinfo{pages}{165} (\bibinfo{year}{2004}).

\bibitem[{\citenamefont{Bennett et~al}(1996)\citenamefont{Bennett, DiVincenzo,
  Smolin, and Wootters}}]{BDSW01a}
\bibinfo{author}{\bibfnamefont{C.~H.} \bibnamefont{Bennett et~al}},
 \bibinfo{journal}{Phys. Rev. A}
  \textbf{\bibinfo{volume}{54}}, \bibinfo{pages}{3824} (\bibinfo{year}{1996}).
  
\bibitem{ChildressComm} L. Childress, J. M. Taylor, A. S. Sorensen, and M. D. Lukin,
Phys. Rev. Lett. {\bf 96}, 070504, 2006.

\bibitem[{\citenamefont{D\"{u}r and Briegel}(2003)}]{DB01a}
\bibinfo{author}{\bibfnamefont{W.}~\bibnamefont{D\"{u}r}} \bibnamefont{and}
  \bibinfo{author}{\bibfnamefont{H.~J.} \bibnamefont{Briegel}},
  \bibinfo{journal}{Phys. Rev. Lett.} \textbf{\bibinfo{volume}{90}},
  \bibinfo{pages}{067901} (\bibinfo{year}{2003}).

\bibitem[{\citenamefont{Duan et~al}(2006)\citenamefont{Duan, Blinov, Moehring, Monroe}}]{duan04}
\bibinfo{author}{\bibfnamefont{L.-M.} \bibnamefont{Duan et~al}},
  \bibinfo{journal}{Quant. Info. and Comp.}
  \textbf{\bibinfo{volume}{4}}, \bibinfo{pages}{165}
  (\bibinfo{year}{2005}).

\bibitem[{\citenamefont{Oi et~al}(2006)\citenamefont{Oi, Devitt, and
  Hollenberg}}]{ODH01a}
\bibinfo{author}{\bibfnamefont{D.~K.~L.} \bibnamefont{Oi}},
  \bibinfo{author}{\bibfnamefont{S.~J.} \bibnamefont{Devitt}},
  \bibnamefont{and} \bibinfo{author}{\bibfnamefont{L.~C.~L.}
  \bibnamefont{Hollenberg}}, \bibinfo{journal}{Physical Review A}
  \textbf{\bibinfo{volume}{74}}, \bibinfo{pages}{052313}
  (\bibinfo{year}{2006}).

\bibitem[{\citenamefont{den Nest et~al}(2007)\citenamefont{den Nest, D\"{u}r,
  Miyake, and Briegel}}]{NDMB01a}
\bibinfo{author}{\bibfnamefont{M.~V.} \bibnamefont{den Nest}},
  \bibinfo{author}{\bibfnamefont{W.}~\bibnamefont{D\"{u}r}},
  \bibinfo{author}{\bibfnamefont{A.}~\bibnamefont{Miyake}}, \bibnamefont{and}
  \bibinfo{author}{\bibfnamefont{H.~J.} \bibnamefont{Briegel}},
  \bibinfo{journal}{New Journal Physics} \textbf{\bibinfo{volume}{9}},
  \bibinfo{pages}{204} (\bibinfo{year}{2007}).
  
   \bibitem[{\citenamefont{Raussendorf et~al}(2003)\citenamefont{Raussendorf,
  Browne, and Briegel}}]{RBB01a}
\bibinfo{author}{\bibfnamefont{R.}~\bibnamefont{Raussendorf}},
  \bibinfo{author}{\bibfnamefont{D.~E.} \bibnamefont{Browne}},
  \bibnamefont{and} \bibinfo{author}{\bibfnamefont{H.~J.}
  \bibnamefont{Briegel}}, \bibinfo{journal}{Phys. Rev. A}
  \textbf{\bibinfo{volume}{68}}, \bibinfo{pages}{022312}
  (\bibinfo{year}{2003}).

\bibitem[{\citenamefont{Hein et~al}(2006)\citenamefont{Hein, D\"{u}r, Eisert,
  Raussendorf, den Nest, and Briegel}}]{HDERNB01a}
\bibinfo{author}{\bibfnamefont{M.}~\bibnamefont{Hein et~al}}, 
 (\bibinfo{year}{2006}), \eprint{quant-ph/0602096}.

\bibitem[{\citenamefont{Raussendorf et~al}(2007)\citenamefont{Raussendorf,
  Harrington, and Goyal}}]{RHG01a}
\bibinfo{author}{\bibfnamefont{R.}~\bibnamefont{Raussendorf}},
  \bibinfo{author}{\bibfnamefont{J.}~\bibnamefont{Harrington}},
  \bibnamefont{and} \bibinfo{author}{\bibfnamefont{K.}~\bibnamefont{Goyal}}
  (\bibinfo{year}{2007}), \eprint{quant-ph/0703143}.

\end{thebibliography}
\end{document}